
\magnification1200
\font\BBig=cmr10 scaled\magstep2
\font\BBBig=cmr10 scaled\magstep3


\def\title{
{\bf\BBBig
\centerline{The Nappi-Witten example}\bigskip \centerline{and}\bigskip
\centerline{gravitational waves}
}}

\def\foot#1{
\footnote{($^{\the\foo}$)}{#1}\advance\foo by 1 } 


\def\authors{
\centerline{C. DUVAL\foot{D\'epartement de Physique, Universit\'e
d'Aix-Marseille II and
Centre de Physique
\hfill\break
Th\'eorique
CNRS-Luminy, Case 907, F-13288 MARSEILLE, Cedex 09 (France)
\hfill\break
e-mail:duval@marcptsu1.univ-mrs.fr.}
Z. HORV\'ATH\foot{Institute for Theoretical Physics, E\"otv\"os University,
 H-1088 BUDAPEST,
\hfill\break
Puskin u. 5-7 (Hungary)
e-mail: zalanh@ludens.elte.hu}
P. A. HORV\'ATHY\foot{D\'epartement de Math\'ematiques, Universit\'e de Tours,
Parc de Grandmont,
\hfill\break
F-37200 TOURS (France) e-mail: horvathy@univ-tours.fr}} }

\def\runningauthors{Duval, Horv\'ath, Horv\'athy}

\def\runningtitle{Strings in WZW\dots}


\voffset = 1cm 
\baselineskip = 14pt 

\headline ={
\ifnum\pageno=1\hfill
\else\ifodd\pageno\hfil\tenit\runningtitle\hfil\tenrm\folio
\else\tenrm\folio\hfil\tenit\runningauthors\hfil \fi
\fi}

\nopagenumbers
\footline = {\hfil} 


\def\and{\qquad\hbox{and}\qquad}

\def\kikezd{\parag\underbar}

\def\IR{{\bf R}}
\def\smallcirc{{\raise 0.5pt \hbox{$\scriptstyle\circ$}}}
\def\smallover#1/#2{\hbox{$\textstyle{#1\over#2}$}}
\def\2{{\smallover 1/2}}

\def\parag{\hfil\break} 
\def\={\!=\!}

\def\IR{{\bf R}}
\def\semidirectproduct{
{\ooalign{\hfil\raise.07ex\hbox{s}\hfil\crcr\mathhexbox20D}} }


\newcount\ch 
\newcount\eq 
\newcount\foo 
\newcount\ref 

\def\chapter#1{
\parag\eq = 1\advance\ch by 1{\bf\the\ch.\enskip#1} }

\def\equation{
\leqno(\the\ch.\the\eq)\global\advance\eq by 1 }

\def\reference{
\parag [\number\ref]\ \advance\ref by 1
}

\ch = 0 
\foo = 1 
\ref = 1 


\title
\vskip 1.5cm
\authors
\vskip .4in

\parag
{\bf Abstract}\hskip 2mm

{\it The vanishing of the anomaly
in the recent example of Nappi and Witten,
constructed from the Wess-Zumino-Witten model
based on a certain non-semisimple group, follows
from a more general result valid for gravitational waves.
The construction of the metric is explained.}

\vskip2cm
\noindent{March 1994}.
\vskip1cm
\noindent{Final Version (0.111)}
\vskip5mm
\noindent{
submitted to the
{\it Phys. Rev. Lett.} as a {\it Comment}} \vfill\eject

Recently, Nappi and Witten [1]
presented a new example of anomaly-free
string propagation on four-dimensional space-time with Lorentz signature.
Their example,
constructed from an ungauged WZW (Wess-Zumino-Witten) model based on a certain
non-semisimple group of the type considered before by Cangemi and Jackiw [2],
has space-time metric and antisymmetric field
$$
d\vec{x}^2-2du\big[dv+\2\epsilon_{ij}x^jdx^i\big]+b\,du^2
\and
b_{ij}=u\,\epsilon_{ij},
\eqno(1)
$$
($i=1,2$) respectively.
Now the metric (1)
is a particular case of Brinkmann's
gravitational pp wave with a
covariantly constant null vector [3],
$$
g_{ij}(\vec{x},u)dx^idx^j-2du\big[dv+A_i(\vec{x},u)dx^i\big] +K(\vec{x},u)du^2.
\eqno(2)
$$

String propagation in a
gravitational wave background has been widely studied [4].
In Ref. [5] --- motivated precisely by the WZW model --- it is shown,
in particular, that for $D=26$ and for the choice $
g_{ij}=g_{ij}(u),
\,
A_i=\2A_{ij}(u)x^j,
\,
b_{iu}=\2B_{ij}(u)x^j,
$
the anomaly vanishes at all loop orders
provided the 1-loop condition
$$
\bigtriangleup K={1\over2}\big(A_{ij}A^{ij}-B_{ij}B^{ij}\big) \eqno(3)
$$
is satisfied.
Then the vanishing of the anomaly in the Nappi-Witten case readily follows
since
the metric (1) corresponds to $g_{ij}=\delta_{ij}$, $A_{ij}=\epsilon_{ij}$ and
$K=b$, and the
Nappi-Witten antisymmetric field is brought by the gauge transformation
$
b_{\mu\nu}\to b_{\mu\nu}+2\partial_{[\mu}\lambda_{\nu]} $
with $\lambda_i=\2u\,\epsilon_{ij}x^j$
to the form (2) with
$B_{ij}=-\epsilon_{ij}=-A_{ij}$.
Eq. (3) is
identically satisfied.

The result can also be
derived from those in Ref. [4]. A
rotation with angle $u/2$ in the plane
carries in fact the metric (1) into
$$
d\vec{x}^2-2dudv+Kdu^2,
\qquad
K=b-\vec{x}^2/4,
\eqno(4)
$$
for which the vanishing of the anomaly at all loop orders has been shown
in Ref. [4] under the sole condition
$
\bigtriangleup K+\2B_{ij}B^{ij}=0,
$
which is clearly the same as our (3). Absorbing the constant $b$
into the \lq vertical' coordinate as $v\to v-b(u/2)$ yields furthermore
an exact plane wave,
$
d\vec{x}^2-2dudv-(1/4)\vec{x}^2du^2,
$
for which the vanishing of anomaly has been proved also non-perturbatively [4].

The symmetries of an
exact plane wave are easy to describe. For example, the three
commuting symmetries noticed by Nappi and Witten can be made
manifest by applying
the coordinate transformation
$
u=\tilde{u},
\,
\vec{x}=\sin(u/2)\,\vec{\tilde{x}},
\,
v=\tilde{v}+(1/8)\sin u\,\vec{\tilde{x}}^2,
$
which brings the exact plane wave metric into the form $
\sin^2(\tilde{u}/2)\,d\vec{\tilde{x}}^2-2d\tilde{u}d\tilde{v}, $
for which the three translations
$\vec{\tilde{x}}\to\vec{\tilde{x}}+\vec{\xi},\, \tilde{v}\to\tilde{v}+\nu$
are obvious isometries [6].
Let us mention that the exact plane wave --- and hence the Nappi-Witten
metric (1) --- is conformal to Minkowski space $d\vec{X}^2-2dUdV$
with conformal factor
$\Omega^2(u)=\cos^{-2}(u/2)$, namely through $
\vec{X}={\vec{x}/\cos(u/2)}
$,
$
U=2\tan(u/2)
$,
$
V=v+(1/4)\vec{x}^2\tan(u/2).
$
\goodbreak
We now explain the arisal of the plane-wave metric on the group.
The group
of isometries of the Euclidian plane,
${\rm{ISO}}(2)={\rm{SO}}(2)\semidirectproduct\IR^2$, is also a
group of canonical transformations for the symplectic structure
$dx\wedge{dy}$. But the $\IR$-bundle $H(1)$ over the plane,
endowed with the $1$-form
$\omega={1\over2}(xdy-ydx)+dv$
is identified with the Heisenberg group. The $\omega$-preserving
transformations of $H(1)$ that cover ${\rm{ISO}}(2)$
form the `diamond'
 ---~or oscillator~--- group
$
G={\rm{SO}}(2)\semidirectproduct H(1),
$
generated by rotations,
$J=y\partial_x-x\partial_y-\partial_u$, translations,
$P_x=\partial_x-{1\over2}y\partial_v$ and
$P_y=\partial_y+{1\over2}x\partial_v$,
and by the central generator
$T=\partial_v$.
Thus $G$ centrally extends ${\rm ISO}(2)$ and is seen
to be precisely the group of Nappi and Witten.

Although non semi-simple, $G$
admits a natural bi-invariant Lorentz metric.
Indeed, the components of the left-invariant Maurer-Cartan
form of $G$ are $\theta=-du$ (rotation),
$\alpha+i\beta=e^{-iu}(dx+idy)$ (translations) and
$\omega$ (extension) as given above.
The combination
$
\alpha^2+\beta^2
+2\theta\omega
+b\theta^2
$
is a right-invariant metric for any constant $b$, and it yields
the metric of Eq.~(1).
Moreover, the null vector
$
T={\partial_v}
$
is covariantly constant.

Note that the Cangemi-Jackiw group [2] arises in
the same manner starting with the Poincar\'e group
${\rm{ISO}}(1,1)$.

\kikezd{Acknowledgment} We are indebted to Professor
R. Jackiw for calling our attention to this problem,
and to Professors P. Forg\'acs and L. Palla for discussions.

\vskip2mm

\centerline{\bf\BBig References}

\reference
C. Nappi and E. Witten, Phys. Rev. Lett. {\bf 71}, 3751 (1993).

\reference
D. Cangemi and R. Jackiw,
Phys. Rev. Lett. {\bf 69}, 233 (1992).

\reference
H. W. Brinkmann,
Math. Ann. {\bf 94}, 119-145 (1925).

\reference
D. Amati and C. Klim\v c\'\i k, Phys. Lett. {\bf B219}, 443 (1989);
R. G\"uven, Phys. Lett. {\bf B191}, 275 (1987);
H. de Vega and N. Sanchez, Nucl. Phys. {\bf B317}, 706 (1989);
A. A. Tseytlin, Phys. Lett. {\bf 288}, 279 (1992);
G. T. Horowitz and A. R. Steif, Phys. Rev. Lett. {\bf 64}, 260 (1990);
Phys. Rev. {\bf D42}, 1950 (1990); A. Steif, Phys. Rev. {\bf D42}, 2150 (1990);
G. T. Horowitz, in Proc. VIth Int. Superstring Workshop {\it Strings'90},
Texas '90, Singapore: World Scientific (1991).

\reference
A. A. Tseytlin, Nucl. Phys. {\bf B390}, 153 (1993);
C. Duval, Z. Horv\'ath and P. A. Horv\'athy, Phys. Lett. {\bf B313}, 10 (1993);
Mod. Phys. Lett. {\bf A8}, 3749 (1993).

\reference
G. W. Gibbons, Commun. Math. Phys. {\bf 45}, 191 (1975).

\bye